\documentclass{ws-procs975x65}

\begin{document}

\title{Degrees of freedom and local Lorentz invariance in $f(T)$ gravity}
\author{Mar\'ia Jos\'e Guzm\'an$^{1,*}$ and Rafael Ferraro$^{2,3}$}

\address{$^1$ Instituto de F\'isica de La Plata (IFLP, CONICET-UNLP), C. C. 67, 1900 La Plata, Argentina\\
$^2$ Instituto de Astronom\'ia y F\'isica del Espacio (IAFE, CONICET-UBA), Casilla de Correo 67, Sucursal 28, 1428 Buenos Aires, Argentina\\
$^3$ Departamento de F\'isica, Facultad de Ciencias Exactas y Naturales, Universidad de Buenos Aires, Ciudad Universitaria, Pabell\'on I, 1428 Buenos Aires, Argentina\\
$^*$E-mail: mjguzman@fisica.unlp.edu.ar}

\begin{abstract}
$f(T)$ gravity is a generalization of the teleparallel equivalent of general relativity (TEGR), where $T$ is the torsion scalar made up of the Weitzenb\"{o}ck connection. This connection describes a spacetime with zero curvature but with nonvanishing torsion, which fully encodes the gravitational phenomena. We will present recent results in $f(T)$ gravity related with the issue of the degrees of freedom of the theory. In particular, we discuss the recent finding that $f(T)$ gravity has one extra degree of freedom compared with TEGR, which was concluded through a detailed Hamiltonian analysis of the constraint structure of the theory. The physical interpretation of this result at the level of the trace of the equations of motion and its comparison with the $f(R)$ case is discussed.
\end{abstract}


\bodymatter

\section{Teleparallel gravity and the $f(T)$ paradigm}
The teleparallel equivalent of general relativity is an alternative formulation of gravity that employs the tetrad field $\mathbf{e}_a = e_a^{\mu} \partial_{\mu}$ as a dynamical variable. Unlike general relativity, which is formulated in a Riemann spacetime equipped with the Levi-Civita connection, teleparallel theories of gravity are commonly defined in terms of the Weitzenb\"{o}ck connection. The Riemann tensor vanishes in this connection, therefore it describes a curvatureless spacetime with absolute parallelism (or \textit{tele}-parallelism). However, the Weitzenb\"{o}ck connection $\Gamma^{\rho}_{\ \mu\nu}= e^{\rho}_a \partial_{\mu} E^a_{\nu}$ has nonvanishing torsion given by $T^{\rho}_{\ \mu\nu} =e_{a}^{\rho} ( \partial_{\mu}E^{a}_{\nu} - \partial_{\nu} E^{a}_{\mu} )$, where $\mathbf{E}^a=E^a_{\mu} dx^{\mu}$ is the co-tetrad field. The action giving dynamical equations that are equivalent to Einstein equations is
\begin{equation}
 S = \dfrac{1}{2\kappa} \int d^4 x\ E \ T = \dfrac{1}{2\kappa} \int d^4 x\ E S_{\rho}^{\ \mu\nu} T^{\rho}_{\ \mu\nu}
\label{Stegr}
 \end{equation}
where $\kappa=8\pi G$, $E=\text{det}(E^a_{\mu})$, and it is implicitly defined the torsion scalar $T$, and the so-called superpotential $S_{\rho}^{\mu\nu}$ is given by
\begin{equation}
 S_{\rho}^{\ \mu\nu} = \dfrac{1}{4}(T_{\rho}^{\ \ \mu\nu} - T^{\mu\nu}_{\ \ \rho} + T^{\nu\mu}_{\ \ \rho} ) + \dfrac{1}{2} T^{\ \sigma\mu}_{\sigma} \delta^{\nu}_{\rho}  - \dfrac{1}{2} T^{\ \sigma\nu}_{\sigma} \delta^{\mu}_{\rho} .
\end{equation}
The equations of motion for the action \eqref{Stegr} are
\begin{equation}
4 e \partial_{\mu}(E e^{\lambda}_a S_{\lambda}^{\ \mu\nu}) + 4 e^{\lambda}_a T^{\rho}_{\ \mu\lambda} S_{\rho}^{\ \mu\nu}  - e^{\nu}_a  T  = - 2 \kappa e^{\sigma}_a \mathcal{T}^{\ \nu}_{\sigma},
\end{equation}
where $\mathcal{T}^{\ \nu}_{\sigma}$ represents the energy-momentum tensor. The equivalence between GR and TEGR theories emerges, both at the levels of the equations of motion and the Lagrangians, when the metric is expressed in terms of the co-tetrad as $g_{\mu\nu}=\eta_{ab}E^a_\mu E^b_\nu$, since it results
\begin{equation}
 R = - T + 2\ e\ \partial_{\mu} (E T^{\mu})
 \label{equiv}
\end{equation}
where $R$ is written in terms of the tetrad field and calculated with the Levi-Civita connection. Therefore, both Lagrangians only differ by a surface term that is integrated out when plugged into the action.

TEGR is a suitable framework for building deformations of GR, since its  Lagrangian has only first order derivatives of the tetrad field, therefore any function of $T$ will give a gravitational theory with second order equations of motion. The first application of modified teleparallel gravities to high energy modifications of GR was through a Born-Infeld approach \cite{Ferraro:2006jd}, where the main purpose was to obtain an early accelerated expansion of the Universe without resorting to an inflaton field. Later there were proposed low energy deformations to GR through the so-called $f(T)$ gravities, intended to be an alternative explanation of the late-time accelerated expansion of the Universe. The $f(T)$ gravity action is given by \cite{Ferraro:2008ey,Bengochea:2008gz,Ferraro:2012wp}
\begin{equation}
 S = \dfrac{1}{2\kappa} \int d^4 x\ E \ f(T),
\end{equation}
and its equations of motion are
\begin{equation}
  4 e \partial_{\mu} (f^{\prime}(T) E e^{\lambda}_a S_{\lambda}^{\ \mu\nu} ) + 4 f^{\prime}(T) e^{\lambda}_a T^{\sigma}_{\ \mu\lambda} S_{\sigma}^{\ \mu\nu} - e^{\nu}_a f(T) = - 2 \kappa e^{\lambda}_a \mathcal{T}^{\ \nu}_{\lambda}.
\label{eomfT}
\end{equation}
These dynamical equations have a very unusual peculiarity: they violate local Lorentz invariance. This happens in a very particular way: given a tetrad $\mathbf{E}^a$ that satisfies the equations \eqref{eomfT}, the local Lorentz transformed tetrad $\mathbf{E}^{a'} = \Lambda^{a'}_{\ a} \mathbf{E}^a $ does not necessarily satisfy them. This is commonly understood as the theory choosing a preferential frame that endows the spacetime with a determined parallelization. Other interpretations of the problem of Lorentz invariance and its relation with the additional degree(s) of freedom of the theory will be reviewed in the next section.

\section{On local Lorentz invariance and Hamiltonian formalism}

The issue about the number and nature of the additional degrees of freedom in modified teleparallel gravity theories is an open question, which has been faced through several approaches. One of the main strategies utilized is the study of the Hamiltonian formulation of $f(T)$ gravity. This has been studied in the past \cite{Li:2011rn}, with the outcome that the theory posesses $n-1$ extra d.o.f. when compared with TEGR. The authors suggested that the extra d.o.f. of $f(T)$ gravity would manifest in a kind of Higgs mechanism, through a massive vectorial field or a scalar field plus a massless vectorial field. So far, it has not been shown such physical equivalence between these hypotetical fields and the $f(T)$ Lagrangian. Moreover, no extra degrees of freedom appears at the level of cosmological perturbations \cite{Li:2011wu,Izumi:2012qj,Chen:2010va,Golovnev:2018wbh}. This strongly suggests that the counting of degrees of freedom in \cite{Li:2011rn} should be revised in the light of new work. Recently the issue about the Hamiltonian formalism of $f(T)$ gravity has been revised by an independent Hamiltonian analysis \cite{Ferraro:2018tpu}, with the outcome that the theory possess only one extra degree of freedom compared with TEGR. This analysis was based in a Hamiltonian formulation of teleparallel gravity \cite{Ferraro:2016wht} built from a Lagrangian quadratic in first-order derivatives in the tetrad field and proportional to a \textit{supermetric} \cite{Ferraro:2016wht} or \textit{constitutive tensor} \cite{Itin:2016nxk}, a Lorentz invariant mathematical object that significantly simplifies the calculation of the Hamiltonian and the Poisson brackets. This formulation shows that the Hamiltonian structure of TEGR is very simple: the traditional Arnowitt-Deser-Misner constraint structure from general relativity has an additional subalgebra  representing local Lorentz transformations that are gauge symmetries of the theory \cite{Ferraro:2016wht}. 

The Hamiltonian formulation of TEGR mentioned before has been used to develop the Hamiltonian formulation of $f(T)$ gravity in the Jordan frame representation of the theory \cite{Ferraro:2018tpu}. This frame is obtained by defining the Legendre transform of the $f(T)$ action with the help of a scalar field, alike to the $f(R)$ gravity case. It is found a novel constraint structure where it appears a new constraint associated with the introduction of the scalar field, which does not commute with a subset of the constraints ${\tilde G}^{(1)}_{ab}$ associated with the Lorentz sector. In particular, the scalar field constraint $G^{(1)}_{\pi}$ has a nonvanishing Poisson bracket  with only one linear combination of the $n(n-1)/2$ Lorentz constraints $\tilde{G}^{(1)}_{ab}$. Both constraints pair up to become second class, while the remaining constraints are first-class. This could be interpreted as a partial violation of Lorentz symmetry in a unique, undetermined combination of boosts and rotations. The counting of degrees of freedom gives that $f(T)$ gravity has $n(n-3)/2+1$ physical d.o.f., this means one additional d.o.f. when compared with TEGR \cite{Ferraro:2018tpu}. 

Other meditations upon the degrees of freedom of $f(T)$ gravity can be found in the literature. For instance, the method of characteristics for hyperbolic partial differential equations has been applied for studying the qualitative behavior of the degrees of freedom \cite{Ong:2013qja}. The authors find evidence for an extra d.o.f., however they assume that the theory would have three extra d.o.f. \cite{Li:2011rn}. The absence of the supposedly missing two extra d.o.f. leads to the authors to conclude that the theory would have superluminal modes and a bad posed Cauchy problem. It is mandatory to revisit these claims in the light of recent research. Other potentially helpful approaches in this respect must be explored, as the proposals of \textit{covariant} approaches where the spin connection is taken different from zero \cite{Krssak:2015oua,Golovnev:2017dox,Hohmann:2018rwf}. Another helpful strategy could be the introduction of Lagrange multipliers that enforce the vanishment of the Riemann tensor, guaranteeing the null curvature condition \cite{Nester:2017wau}. Finally, considerations about remnant symmetries on the theory \cite{Ferraro:2014owa} and the search for solutions to $f(T)$ through a null tetrad approach \cite{Bejarano:2014bca,Bejarano:2017akj} should be further addressed.

In what follows, we will explore one possible approach for understanding the nature of the extra d.o.f. in $f(T)$ gravity, through its comparison with the well known $f(R)$ gravity case. We will find convincing evidence for its existence in the trace of the equations of motion, however it is noteworthy to point out that the interpretation of the extra d.o.f. in both theories is radically different.

\section{The nature of the extra d.o.f.}

One of the most studied modifications of general relativity is $f(R)$ gravity. The action of this gravitational theory is given by 
\begin{equation}
S_{f(R)} = \dfrac{1}{2\kappa} \int d^4 x\ \sqrt{-g} f(R) + S_m(g_{\mu\nu}),
\end{equation}
where $S_m$ is the action for matter minimally coupled to gravity. This action has fourth-order dynamical equations given by
\begin{equation}
f'(R) R_{\mu\nu} - \dfrac{1}{2} f(R) g_{\mu\nu} - [ \nabla_{\mu} \nabla_{\nu} - g_{\mu\nu} \Box ] f'(R) = \kappa T_{\mu\nu}.
\label{eomfR}
\end{equation}
By taking the trace of the equations of motion it is obtained a relation between the Ricci scalar $R$ and the trace $\mathcal{T}$ of the energy-momentum tensor that is algebraical, that is
\begin{equation}
f'(R) R - 2 f(R) + 3 \Box f'(R) = \kappa \mathcal{T}.
\label{tracefR}
\end{equation}
This equation shows the propagation of a new degree of freedom related with $f'(R)$ \cite{Olmo:2006eh,Sotiriou:2008rp}. Moreover, it is possible to write \eqref{eomfR} and \eqref{tracefR} as second-order equations for the metric and the scalar object $f'(R)$ by changing the notation to
\begin{equation}
 \phi \equiv f'(R), \ \ \ \ V(\phi) \equiv R \phi - f(R),
 \label{LegTr}
\end{equation}
to rewrite Eq.\eqref{eomfR} as
\begin{equation}
 R_{\mu\nu} - \dfrac{1}{2} g_{\mu\nu} R = \dfrac{\kappa}{\phi} \mathcal{T}_{\mu\nu} - \dfrac{g_{\mu\nu}}{2\phi} V(\phi) + \dfrac{1}{\phi}[\nabla_{\mu}\nabla_{\nu} \phi - g_{\mu\nu} \Box \phi ].
 \label{eomfRJF}
\end{equation}
We recognize in Eq.\eqref{LegTr} the Legendre transform of $f(R)$, hence it is also obtained that $R=V^{\prime}(\phi)$, which together with \eqref{eomfRJF} are the dynamical equations associated with the action
\begin{equation}
 S_{JF}[g_{\mu\nu}, \phi] = - \dfrac{1}{2\kappa} \int d^4 x \sqrt{-g} [\phi R - V(\phi) ] + S_{\text{matter}}.
 \label{LagrJF}
\end{equation}
For completing the Legendre transformations it is needed that $\phi$ be a function of the scalar curvature $R$ and moreover, that the Lagrangian in \eqref{LagrJF} is equivalent to the Legendre transform of the function $V(\phi)$, therefore it can be rewritten as a function of $R$ by defining $f(R)=\phi R - V(\phi)$. The trace of Eq. \eqref{eomfRJF} can be regarded as a wave equation for a self-interacting scalar field $\phi$, since it satisfies
\begin{equation}
 3 \Box \phi - \phi^3 [\phi^{-2}V(\phi) ]^{\prime} = \kappa \mathcal{T}.
 \label{extradoffR}
\end{equation}
In this way, $\phi$ looks as a scalar field minimally coupled to the metric $g_{\mu\nu}$, but also coupled to the matter through the trace of the energy-momentum tensor. Meanwhile, Eq.\eqref{eomfRJF} are Einstein equations, on which the source are the matter and the scalar field. The strategy used through the redefinition \eqref{LegTr} allows to rephrase $f(R)$ gravity as a scalar-tensor theory with second-order dynamical equations. The equations \eqref{eomfRJF} describe the dynamics of the metric field, therefore they encompass two d.o.f. in four dimensions. Additionally, there is one extra d.o.f. described by \eqref{extradoffR}.

A reasonable approach would be to perform the same procedure to the equations of motion of $f(T)$ gravity, that is to say, to develop an analogous calculation of the trace of the equations of motion. The analogy is not complete, due to the fact that, while the equations of motion of $f(R)$ gravity are second-order on $f'(R)$, the dynamical equations for $f(T)$ gravity are first-order in the analog variable $f'(T)$. We can clearly see this feature if we rewrite the equations of motion \eqref{eomfT} through the Legendre transform
\begin{equation}
 \phi = f'(T), \ \ \ \ V(\phi)=T \phi - f(T),
\end{equation}
to become
\begin{equation}
4 \phi^{-1} e \partial_{\mu} (\phi E e^{\lambda}_a S_{\lambda}^{\ \mu\nu} ) + 4 e^{\lambda}_a T^{\sigma}_{\ \mu\lambda} S_{\sigma}^{\ \mu\nu} - e^{\nu}_a T = - \dfrac{2\kappa}{\phi} e^{\lambda}_a \mathcal{T}_{\lambda}^{\ \nu} + e^{\nu}_a \dfrac{V(\phi)}{\phi}.
\label{eqfTscalar}
\end{equation}
These dynamical equations keep the structure of the TEGR equations, except for the renormalization of the volume $E$, the gravitational constant $\kappa$, and an additional term $\phi^{-1} V(\phi)$ that behaves as a local cosmological constant. Then the trace of \eqref{eqfTscalar} easily writes as
\begin{equation}
2 T^{\mu} \partial_{\mu}\phi + 2 \phi e \partial_{\mu}( E T^{\mu}) - T\phi + 2V(\phi) = - \kappa \mathcal{T},
\end{equation}
or alternatively as
\begin{equation}
 2 T^{\mu} \partial_{\mu} \phi + 2 V(\phi) + \phi R = - \kappa \mathcal{T},
 \label{TrfTJF}
\end{equation}
when substituting $2 e \partial_{\mu}(E T^{\mu}) $ by $R+T$, as stated in Eq.\eqref{equiv}. We return to the TEGR case when $\phi=1$ and $V(\phi)=0$; in this case the equation \eqref{TrfTJF} implies $R=-\kappa \mathcal{T}$, the classical result obtained in GR. Therefore, it is fair to say that \eqref{TrfTJF} encodes more information than merely GR, and could be interpreted as the description of the propagation of an extra scalar degree of freedom that was not present in TEGR. It has been suggested that the existence of an extra d.o.f. in $f(T)$ is connected with the loss of a gauge symmetry. This can be seen in terms of the symmetries of $T$, which do not remain invariant under general local Lorentz transformations. Henceforth, $f(T)$ will inherit a remnant gauge symmetry, which has an on-shell character \cite{Ferraro2018b}. The implications of this remnant symmetry and its relevance in cosmological solutions deserves further research.

\section{Conclusions and future work}

In this work we have introduced teleparallel gravity and its simplest modification, the so-called $f(T)$ gravity, and briefly reviewed the important issue about the degrees of freedom in this theory. We have discussed several approaches on the counting of d.o.f., in particular some results obtained through the Dirac-Bergmann algorithm for constrained Hamiltonian systems. Recent work suggests that $f(T)$ gravity would have only one extra d.o.f., in oposition to previous research. These claims are checked through the comparison with the $f(R)$ gravity case. In both theories the calculation of the trace of the equations of motion reveals an additional scalar d.o.f.. Nonetheless, its interpretation in the $f(T)$ case is qualitatively different, as it satisfies a first-order differential equation. This equation shows evidence that the additional scalar field could be related to the proper parallelization of spacetime. It is strongly encouraged further research about the characterization of the d.o.f. of more general teleparallel gravities and its implications in cosmology and compact objects.

\end{document}